\documentclass[11pt]{article}
\usepackage{hyperref}
\pdfoutput=1

\begin{document}
\title{Flow Control with Dynamic Roughness}
\author{Vinay Jakkali, Wade W. Huebsch,\\ Patrick H. Browning, and Shanti D. Hamburg
\\ \\ Mechanical and Aerospace Engineering,
\\ \\ West Virginia University, Morgantown, WV 26506, USA}
\maketitle

\begin{abstract}
This paper discusses a fluid dynamics video showing the use of Dynamic Roughness to reattach upper surface flow on a stalled NACA 0012-based wing.
\end{abstract}

\section{Introduction}

Dynamic Roughness (DR) is an active method of flow control currently under investigation by researchers at West Virginia University (WVU) \cite{Jakkali_2013, GallPhD_2010, Huebsch_2012}.
In the submitted video, smoke flow visualization (laser-illuminated atomized olive oil) techniques are employed to demonstrate the efficacy of DR on a NACA 0012-based wing. The wing had a span of 3 inches and a chord length of 4 inches, and was tested with a 6 inch diameter circular end plate in a 6 inch x 6 inch low turbulence Eiffel wind tunnel at WVU. The 2D tests were conducted at chord-based Reynolds numbers of 25,000 and 50,000 – results for the {$ Re_{c}$} = 25,000 case are provided in the video. A wing with no DR element array installed (henceforth referred to as the “clean” case) was first tested to generate a clear understanding of low Reynolds number flow over the wing at an angle of attack of 13$^{\circ}$. All subsequent tests (clean and DR cases) were tested at the same 13$^{\circ}$ angle of attack. It is evident in the video that flow separation on the clean case initiates at the leading edge of the wing. DR is an active method of flow control utilizing compliant surface elements that move from an initial position flush with the wing’s surface upward into, but not leaving, the local boundary layer. An animation is shown in the video to help viewers understand DR motion, and a brief video clip also shows DR elements under actuation from a zoomed-in perspective. The video continues by simultaneously comparing several different cases of clean and DR operation with particular emphasis on the effect of varying DR actuation frequency. It is clear from the first comparison that there is a distinct difference between clean and DR actuated cases for general upper surface flow attachment: DR appears to reattach the otherwise separated flow. It is also apparent, as shown in subsequent comparison videos, that there exists a minimum threshold frequency, {\it $f_{T}$}, for which DR has no apparent effect on upper surface flow. In the Reynolds number case tested, the threshold frequency for the NACA 0012-based 2D wing operating at 13$^{\circ}$ angle of attack is {\it $f_{T}$} {$ \simeq$} 23 Hz. Operating at frequencies above {\it $f_{T}$}, DR is shown to have intermittent efficacy in flow reattachment that tends to improve as {\it f} goes from 40 Hz to a maximum value of 87 Hz (limited by the maximum speed of the cyclic positive displacement DR drive actuation system).


\end{document}